\documentclass[aps,pra,reprint,groupedaddress, amsmath,amssymb]{revtex4-2}
\usepackage{graphicx} 
\usepackage[utf8]{inputenc}
\usepackage{textcomp}
\usepackage[german, english]{babel}
\usepackage{float}
\usepackage{braket}
\usepackage{tikz}
\usepackage{bm}
\usepackage[ruled]{algorithm2e}
\usepackage{xcolor}
\usepackage[nolist]{acronym} 
\usepackage{enumitem}
\usepackage{hyperref}
\usepackage[format = hang]{caption}
\usepackage{subcaption}
\usepackage{floatpag}
\usepackage{afterpage}

\usepackage[section]{placeins}

\usetikzlibrary{shapes,arrows}
\usetikzlibrary{positioning}

\addto\extrasgerman{

}
\addto\extrasenglish{

}

\tikzstyle{decision} = [diamond, draw, fill=blue!20, 
    text width=4.5em, text badly centered,
     inner sep=0pt]
\tikzstyle{block} = [rectangle, draw, fill=blue!20, 
    text width=5em, text centered, rounded corners, minimum height=4em]
\tikzstyle{line} = [draw, -latex',line width = 0.5mm]
\tikzstyle{cloud} = [draw, ellipse,fill=red!20,
    minimum height=2em]

\makeatletter
\AtBeginDocument{%
  \renewcommand*{\AC@hyperlink}[2]{%
    \begingroup
      \hypersetup{hidelinks}%
      \hyperlink{#1}{#2}%
    \endgroup
  }%
}
\makeatother

\DontPrintSemicolon
\SetStartEndCondition{ }{}{}%
\SetKwProg{Fn}{def}{\string:}{}
\SetKwFunction{Range}{range}
\SetKw{KwTo}{in}\SetKwFor{For}{for}{\string:}{}%
\SetKwIF{If}{ElseIf}{Else}{if}{:}{elif}{else:}{}%
\SetKwFor{While}{while}{:}{fintq}%
\AlgoDontDisplayBlockMarkers\SetAlgoNoEnd\SetAlgoNoLine%
\SetKwInOut{Input}{Input}
\SetKwInOut{Output}{Output}
\newtheorem{definition}{Definition}
\newtheorem{algorithmdescription}{Algorithm}

\begin{document}

\title[]{Symmetries for a High Level Neural Decoder on the Toric Code}
\author{Thomas Wagner}
\email{thomas.wagner@uni-duesseldorf.de}
\author{Hermann Kampermann}
\author{Dagmar Bru\ss}
\affiliation{Institute of Theoretical Physics III, Heinrich-Heine-Universität Düsseldorf}

\date{\today}

\begin{abstract}
Surface codes are a promising method of quantum error correction and the basis of many proposed quantum computation implementations. However, their efficient decoding is still not fully explored. Recently, approaches based on machine learning techniques have been proposed by \textcite{NeuralDecoder} as well as \textcite{Varsamopoulos_2017}. In these approaches, a so called high level decoder is used to post-correct an underlying decoder by correcting logical errors. A significant problem is that these methods require large amounts of training data even for relatively small code distances. The above-mentioned methods were tested on the rotated surface code which encodes one logical qubit. Here, we show that they are viable even for the toric surface code which encodes two logical qubits. Furthermore, we explain how symmetries of the toric code can be exploited to reduce the amount of training data that is required to obtain good decoding results. Finally, we compare different underlying decoders and show that the accuracy of high level decoding noticeably depends on the quality of the underlying decoder in the realistic case of imperfect training.
\end{abstract}

\maketitle

\section{Introduction}
A great challenge in the practical realization of quantum computing is the presence of noise which spoils accurate control of physical systems. The effect of such noise can be mitigated by using quantum error correction. The physical state of a system is encoded  into the logical state of a quantum code. Then, computations can be performed on the logical level of the code. As coding introduces redundancy in the data, many errors can be detected and corrected by a decoder. According to the threshold theorem \cite{ThresholdTheorem1,ThresholdTheorem2}, quantum error correction allows us to perform quantum computations with arbitrary accuracy as long as all single component error rates are below a certain threshold. A promising approach to quantum error correction is the use of topological quantum codes. The surface code by \textcite{KitaevSurfaceCode}\cite{KitaevTC} possesses a high threshold error rate \cite{ThresholdOfTC} above some existing experimental error rates \cite{DiCarlo2009}. Furthermore, it has the advantage of only requiring nearest neighbour interactions.  However, a problem in the practical realization of surface codes is the need for decoders that are both fast and accurate. Fast decoding is crucial because the decoding procedure should be shorter than the coherence time of the qubits, which can be of order 1 $\mu s$ for superconducting qubit architectures \cite{DiCarlo2009}. While higher coherence times of order 10 $s$ can be reached in ion trap qubits \cite{HartyIonTraps}, superconducting qubits are currently the main candidate for experimental surface code realizations.

Several different decoders based on various approximations  have been proposed  \cite{DecodingBravyi, DecodingDuclosCianci, DecodingHutter}. These decoders are generally based on the assumption of independent Pauli noise, and it is not always clear how they can be adapted to experimental noise.
Recently, there has been an increasing interest in decoders based on machine learning techniques. These decoders are trained on a set of known errors and then learn to generalize to unknown errors. It is expected that such decoders can adapt to experimental noise, e.g. it has been demonstrated that they can adapt to different rates of stochastic Pauli errors \cite{Varsamopoulos_DesigningNNBasedDecoders}. The first such decoder was developed by \textcite{NeuralDecoder} and is based on stochastic neural networks. It was introduced for the toric  surface code with only phase-flip errors, but the techniques are generalizable to all stabilizer codes. Another approach, called \acl{hld}, based on more conventional \aclp{ffnn} was proposed by \textcite{Varsamopoulos_2017}\cite{Varsamopoulos_DesigningNNBasedDecoders} and further explored by \textcite{Chamberland_2018}. This approach was implemented on the so called rotated surface code, which encodes one qubit, for different noise models including circuit noise. In \cite{Chamberland_2018} it is concluded that, once the decoder is trained, the actual decoding procedure of \acl{ffnn} based decoders is fast enough to be scalable to larger codes. A high performance computing platform is still required. Furthermore, \textcite{MaskaraVersatileNNDecoder} demonstrated that the method is applicable to different architectures, such as color codes and toric codes on triangular lattices, and various noise models. However, it is also pointed out in \cite{Chamberland_2018} that the training of decoders becomes increasingly difficult for larger codes. So far, the method could only be demonstrated for small codes with a distance less than seven. The amount of training data needed to train the networks for larger codes is infeasible. One way to approach this problem are decoders based on local regions of the code \cite{VarsamopoulosDistributedDecoders}\cite{XiaotongNNDecodersForLargeToricCodes}. This technique is inspired by the renormalization group decoder \cite{DecodingDuclosCianci}.

To supplement these approaches, in this paper, it will be shown how symmetries of the toric code can be explicitly incorporated into the training of (feed forward) neural network based decoders. This reduces the amount of training data needed substantially and improves the quality of training. Our approach will be demonstrated for the high level decoder developed by \textcite{Varsamopoulos_2017}\cite{Varsamopoulos_DesigningNNBasedDecoders}, but it is applicable to general machine learning based decoders. This decoder was chosen as an example because it is a relatively simple but still effective machine learning based decoder, and because it was well explored in previous work  \cite{Varsamopoulos_2017,Varsamopoulos_DesigningNNBasedDecoders,Chamberland_2018}.
Furthermore, it is demonstrated that it is possible to train good high level decoders for the toric code encoding two logical qubits. Previous literature considered the rotated surface code which only encodes one logical qubit. The main difference here is in the number of possible logical errors, which is larger by a factor of four for the toric code.

This paper is structured as follows. First, a short introduction about the toric code is given, and standard decoders for the code are reviewed. Next, the high level decoder scheme \cite{Varsamopoulos_2017} is described. Then, it will be explained how symmetries of the toric code can be exploited to improve this decoder. Finally, some numerical results will be presented which demonstrate the increase in performance provided by the inclusion of symmetries.

\section{The Toric Code and Noise Model}

Although the core ideas are applicable to a wider range of codes, the techniques in this paper will be constructed for the toric code developed by \textcite{KitaevSurfaceCode}\cite{KitaevTC}. We give a short description of this code here. (See \cite{SCQCReview} for an in depth review.) For this, the concept of the Pauli operators will be needed:

\begin{definition}[Pauli Group]
The 4 Pauli operators acting on 1 qubit are given in the standard basis by:
\begin{equation}
I = \left( \begin{array}{rr}1 & 0 \\0 & 1\\ \end{array}\right)
X = \left( \begin{array}{rr}0 & 1 \\1 & 0\\ \end{array}\right)
Y = \left( \begin{array}{rr}0 & i \\-i & 0\\ \end{array}\right)
Z = \left( \begin{array}{rr}1 & 0 \\0 & -1\\ \end{array}\right)
\end{equation}
The Pauli group on $n$ qubits is the multiplicative group generated by all possible n-fold tensor products of Pauli operators and the imaginary unit i.

\end{definition}

\begin{figure}
\centering
\includegraphics[width = 0.5\textwidth]{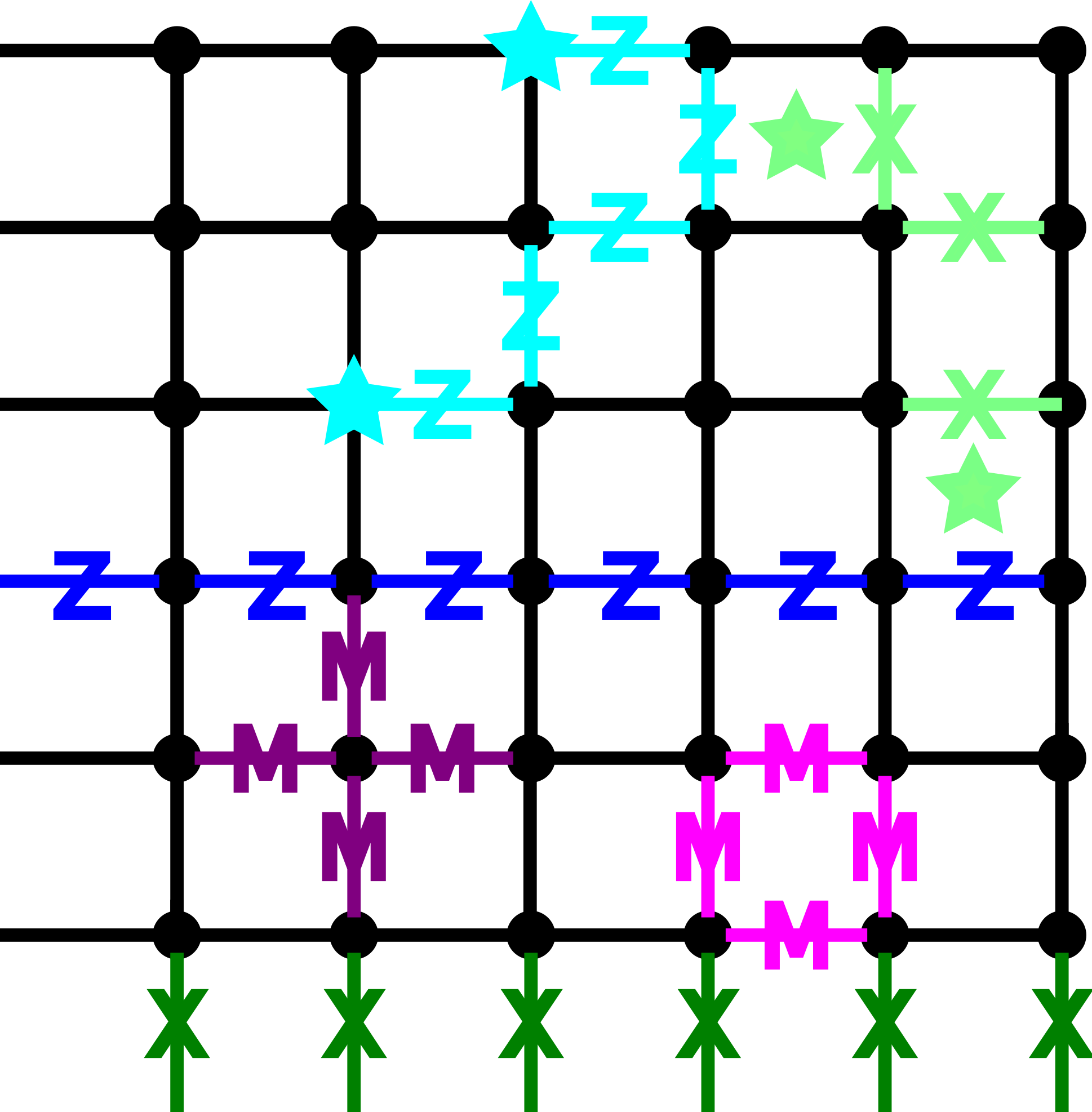}
\caption{Representation of the $6 \times 6$ toric code. Note that the boundary of the lattice is periodic. The edges leaving at the left border wrap back around to the right and the edges leaving at the bottom wrap back around to the top. Examples of different error chains are shown. Marked with $Z$, a logical $Z_2$ operator is shown in dark blue in the middle and a detectable $Z$ error chain is shown in bright blue at the top. Its syndrome is marked by the bright blue stars on the corresponding vertices. Similarly, marked with $X$, a logical $X_1$ operator is shown in dark green at the bottom and a detectable $X$ error chain is shown in bright green at the top. Marked with $M$, a star operator is shown in dark purple on the left, and a plaquette operator is shown in bright pink on the right.}
\label{TCExample}
\end{figure}

 The toric code is a stabilizer code that is defined on an $L\times L$ square lattice embedded on a torus. The lattice consists of vertices, edges and faces. Four edges are connected to each vertex, and each face is surrounded by four edges (\autoref{TCExample}). Each edge of the lattice is associated with a physical qubit of the code. In the following we denote the Pauli X / Pauli Z operator acting on the qubit associated with edge $e$ of the lattice by $X_e$ / $Z_e$. The vertices and faces of the lattice are associated with stabilizer operators. Each vertex $v$ represents a \textit{star} operator:
\begin{equation}
X_v = \prod\limits_{e \in \partial^0 v} X_e
\end{equation}
where $\partial^0 v$ is the coboundary of $v$, i.e. the set of four edges connected to $v$.  Similarly, each face $f$ represents a \textit{plaquette} operator:
\begin{equation}
Z_f = \prod\limits_{e \in \partial_1 f} Z_e
\end{equation}
where $\partial_1 f$ is the boundary of $f$, i.e. the set of four edges adjacent to $f$. The stabilizer group $S$ consists of all possible products of the stabilizer operators above. The toric code is then defined as the common eigenspace with eigenvalue +1 of all operators in $S$. The code encodes two logical qubits (i.e. the code space is 4-dimensional), because $S$ is generated by $2L^2 - 2$ independent generators and there are $2L^2$ physical qubits in the code. Pauli operators acting on the code can be represented as chains of edges on the lattice, by marking which qubits are affected by a Pauli $Z$ or Pauli $X$ operator. We call a Pauli operator a \textit{logical operator} if it maps states in the code space back into the code space. We consider two logical operators equivalent if they only differ by an element of the stabilizer group. Up to this equivalence, there are $16$ logical operators, corresponding to the two qubit Pauli group. These operators correspond to loops on the lattice, as illustrated in \autoref{TCExample}. They will be denoted with a subscript to indicate which logical qubit they act on, e.g. the logical $Z_1$ operator is the logical $Z$ operator that acts on the first logical qubit.

We consider the effect of Pauli errors on the code. If no errors affect the code, the measurement of each stabilizer operator will result in a $+1$ outcome. If an error $e$ affects the code, it might anti-commute with some stabilizer operators. The stabilizers that anti-commute with the error will flip their measurement outcome to $-1$. These are called \textit{detections}, and together they form the \textit{syndrome} of the error. As an example, in the case of a $Z$ error chain, the detections are located at the vertices at the end points of the error chain on the lattice. For these errors the task of a decoder is to propose, based on the syndrome, a recovery chain $r$ that eliminates the error. The recovery was successful if the product of the error and the recovery lies in the stabilizer group. As an example, for $Z$ errors this means that the error and the recovery form the boundary of a region on the lattice. This also implies that all recoveries that only  differ by stabilizer operators are logically equivalent. It is therefore not necessary to deduce the exact error that occurred, but only its equivalence class up to stabilizer applications.

We will consider the toric code subject to local depolarizing noise. In this model, errors occur independently on each physical qubit. Each physical qubit is either unaffected by noise with probability $1-q$ or replaced by the completely mixed state with probability $q$. The action on the density operator $\rho$ of one qubit is therefore expressed by the quantum channel:
\begin{equation}
\rho \mapsto (1-q)\rho + q\frac{I}{2} = (1 - \frac{3}{4}q)\rho + \frac{q}{4}(X\rho X + Y\rho Y + Z\rho Z) 
\label{eq:DepolarizingChannel}
\end{equation}
Thus, the channel can be simulated by leaving each qubit untouched with probability $1 - p$, where  $p = \frac{3}{4}q$, or applying exactly one of the three Pauli operators each with probability $\frac{p}{3}$. The error rate $p$ will also be referred to as \textit{depolarizing noise parameter}.
Note that while the stabilizer measurements are assumed to be perfect here for simplicity, the methods presented in this paper do not depend on this assumption. The pre-processing presented in \autoref{sec:Symmetries} can also be used with imperfect syndrome measurements as long as the symmetries of the code are not broken, i.e. measurement errors must be equally likely for all stabilizers. The effect of imperfect syndrome measurements on standard \aclp{hld} has been explored in \cite{Varsamopoulos_DesigningNNBasedDecoders}.

\section{Simple Decoders for the Toric Code}
\label{sec:SimpleDecoders}
Here, we will shortly describe two simple ways of decoding the toric code.

The first is \acfi{mwpm} \acused{mwpm}based on Edmonds Blossom algorithm \cite{Edmonds-Blossom}. This decoder will be used as a benchmark throughout this paper. Here, $Z$ and $X$ errors are decoded independently. The $Z$  / $X$ recovery is found by proposing the shortest chain that matches the syndrome of the vertices / faces. This corresponds to finding the lowest weight error matching the syndrome, i.e. the error acting on as few physical qubits as possible. In our paper an implementation based on the \textit{NetworkX} python package \cite{NetworkX} is used. There are two problems with \ac{mwpm} decoding. The first is that because $Z$ and $X$ errors are decoded independently, $Y$ errors can lead to incorrect decoding. Essentially, a $Y$ error is counted as two separate errors which is only correct if $X$ and $Z$ errors are independent. In the depolarizing noise model this assumption is not correct. An example of this problem can be found in \cite{DecodingDuclosCianci}. The second problem is that \ac{mwpm} does not account properly for the effect of degeneracy. All errors that only differ by a stabilizer operator are logically equivalent. Therefore, it can happen that the most likely class of equivalent errors does not contain the most likely (shortest) error. This leads to suboptimal decoding. An example can again be found in \cite{DecodingDuclosCianci}. The runtime of (unoptimized) \ac{mwpm} scales as $\mathcal O(L^6)$ \cite{DecodingDuclosCianci}, which is already a problem for larger codes.

Therefore, it will be useful to introduce a simpler decoder. This \textit{trivial decoder} is designed to return a recovery as fast as possible. First, we enumerate the stabilizer operators in some way, say from top left to bottom right in the lattice picture. The trivial decoder then works by matching the detections in a syndrome iteratively according to the above enumeration, using the shortest chain for each matching. This means the first detection is connected with the second, the third with the fourth and so on. Because the measurements are assumed to be perfect the total number of detections will always be even, so no detections are left unmatched. Because the number of expected detections increases quadratically in $L$, the runtime of this algorithm will also be quadratic in $L$. The recovery proposed by this decoder is very inaccurate, but it will be useful as an initial decoding after which we apply a so called \acfi{hld}.

\section{High Level Neural Decoder}

\begin{figure}[t]
\centering
%
\includegraphics[width=\linewidth]{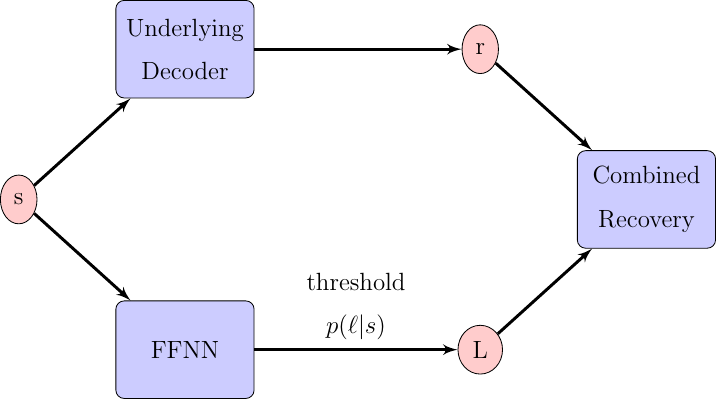}
\caption{The decoding process of the \acl{hld}. A syndrome $s$ is decoded by the underlying decoder to obtain a physical recovery $r$, and by an \ac{ffnn} to obtain the logical error $\ell$ of the underlying decoder. The logical error is applied as post-correction to the physical recovery to obtain a combined recovery.}
\label{fig:HLD-Decoding}
\end{figure}

In \cite{Varsamopoulos_2017} it was shown how decoding can be approached as a classification problem, which is standard in machine learning. Given a syndrome on the toric code, first some standard decoder is used which proposes a recovery chain that matches the measured syndrome. This will be refered to as the \textit{underlying decoder}. Because the proposed recovery matches the syndrome, the product of the error and the recovery will form a logical operator. The classification task is then to predict the most likely logical operator based on the initial syndrome that was measured. Then, an additional recovery corresponding to the predicted logical operator can be applied. This essentially constitutes a post-correction of the underlying decoder. The basic problem is to correctly classify input vectors, corresponding to syndroms, into different classes, corresponding to logical operators. This decoding scheme is called a \acfi{hld}\acused{hld}. In \cite{Varsamopoulos_2017} a surface code which encodes one logical qubit, called the rotated surface code, was considered. Therefore there were 4 possible logical errors, corresponding to a classification problem with 4 classes. Here we will consider the toric code, which encodes two qubits. Therefore the classification problem has 16 classes, corresponding to the two qubit Pauli group. The decoding process is illustrated schematically in \autoref{fig:HLD-Decoding}, using \ac{mwpm} as the underlying decoder.

The classification task outlined above is approached with a simple and widely used machine learning model known as \acfi{ffnn}. An \ac{ffnn} consists of several layers of real-valued units. The first layer corresponds to the input vector, and the last layer has one unit for each possible class label. Between them are several hidden layers. Each hidden layer applies a transformation of the form $\bm y = g( W \bm x + \bm b)$ to the values of the previous layer, where the matrix $W$ and the vector $\bm b$ are free parameters that will be learned during training. They are called the weights and biases of the layer. The weights and biases of all layers together form the parameter vector $\bm \theta$ of the network. The function $g$ is called the activation function. In this work it is chosen to be the rectified linear unit:
\begin{equation}
g(x) = \max(0,x) 
\end{equation}
applied elementwise to the vector. This is a standard choice in machine learning, for example suggested in \cite{Goodfellow}. The output layer instead uses the softmax activation function:
\begin{equation}
\textrm{softmax}(\bm x)_i = \frac{\exp(x_i)}{\sum_j \exp(x_j)}
\end{equation}
which is necessary for classification tasks. The parameters $\bm \theta$ of the model are found by considering a training set $T = \{(e,\ell)\}$ of errors with known logical errors, generated according to the depolarizing noise model. This training set defines a cross-entropy loss function:
\begin{equation}
E(\bm \theta) = \sum_{(e,\ell) \in T} \ln(y_{\ell}(\bm x;\bm \theta))
\end{equation}
where $y_{\ell}$ is the component of the output layer corresponding to the logical error $\ell$. This loss function can be further modified by adding a \textit{weight decay} term $\lambda \|\bm \theta\|_2$ for some positive $\lambda$. This can help against overfitting issues by keeping the network parameters smaller \cite{Goodfellow}. The parameters $\bm \theta$ are found by minimizing this loss function with the adaptive moment estimation algorithm \cite{Adam}, which is a variant of stochastic gradient descent. Before the first iteration of stochastic gradient descent the parameters are initialized randomly from a normal distribution. After the training, the model can be evaluated on test sets that were generated independently from the training set. Sometimes, we will also consider the logical error rate of the \acl{hld} on the training set itself. We refer to this as the "training error". All these methods were implemented with the Shark machine learning library \cite{Shark}.

The model has a number of hyperparameters that need to be chosen in advance. These are: The number $n_{it}$ of iterations of stochastic gradient descent, the learning rate $\eta$ used in stochastic gradient descent, the number $n_h$ of hidden layers, the numbers of units $l_i$ in the $i$'th hidden layer, the strength $\lambda$ of weight decay regularization, and the width of the distribution used for initialization. These parameters need to be chosen sensibly according to some heuristic, usually with some trial and error involved.

It should be stressed that the accuracy of the model strongly depends on the quality and size of the training set. If the training set is too small the model will be unable to learn the real distribution of logical errors. This usually manifests in overfitting, i.e. the accuracy on the training set is good but the accuracy on the test sets is bad. This is especially problematic for larger code distances, where a large amount of different syndroms should be represented in the training data. Finally, it should be noted that the best performance is reached if the training set is generated at the same physical error rate as the test set the model should be used for \cite{Varsamopoulos_DesigningNNBasedDecoders}. However, the models can still generalize well to different error rates. 

In this paper, we restrict ourselves to the simplest version of the high level decoder by using only feed forward neural networks. This version requires relatively little hyperparameter tuning, and is therefore well suited to exploring the effect of the new techniques introduced in the next sections. It should however be noted that more sophisticated network architectures, like recurrent neural networks or convolutional neural networks, can yield better decoding accuracy, especially if one considers error models with imperfect syndrome measurements \cite{Chamberland_2018, Varsamopoulos_DesigningNNBasedDecoders}.

\section{Symmetries of the Toric Code}
\label{sec:Symmetries}

In order to learn the correct conditional distributions of logical errors given syndromes, the model needs to have a large selection of syndromes available in the training data. Preferably each syndrome should appear multiple times to make the prediction of the most likely logical error more accurate. For a $7 \times 7$ surface code the input space consists of $2^{98}$ different possible syndromes so the amount of training data needed is already very large. Here, we will describe how symmetries of the code can be explicitly incorporated into the training of decoders in order to reduce the effective size of the input space. This reduces the amount of training data that is needed or, alternatively, allows for better results with the same amount of training data.

There are several symmetries on the toric code, including exchange, translation and mirror symmetry. Here we will focus mainly on translation and exchange symmetry. Translation symmetry means that the code is invariant under a translation of the vertices, edges and plaquettes, taking into account the periodic boundary conditions:
\begin{definition}[Translation]
The translation of an $L \times L$ lattice by $a$ steps to the left and $b$ steps to the top is obtained by mapping the vertex at position $(x,y)$ in the lattice to the vertex at position $(x-a \textrm{ \textnormal{mod} } L,y-a \textrm{ \textnormal{mod} } L)$ in the lattice, and analogously mapping edges and plaquettes.
\end{definition}

 As an example, the syndrome shown in \autoref{fig:TranspositionAndTranslationCompatibility2} is obtained by translating the syndrome shown in \autoref{fig:TranspositionAndTranslationCompatibility1} one step to the left and one step to the top.

Exchange symmetry means that the toric code is invariant under an exchange of the toroidal and poloidal direction on the torus, provided one chooses a lattice with the same number of edges in both directions. The exchange does however correspond to a relabeling of the logical operators. In the lattice view of the code, which is shown in \autoref{TCExample}, this corresponds to an anti-transposition of the lattice, i.e. reflecting over the anti-diagonal:
\begin{definition}[Anti-Transposition]
The anti-transposition of an $L \times L$ lattice is obtained by mapping the vertex at position $(x,y)$ in the lattice to the vertex at position $(L-y+1,L-x+1)$, and analogously for edges and plaquettes.
\end{definition}

As an example, the syndromes shown in \autoref{fig:TranspositionAndTranslationCompatibility3} and \autoref{fig:TranspositionAndTranslationCompatibility1} differ by such an anti-transposition. Note that if one considers surface codes that include holes or different boundary conditions, the symmetries mentioned above might be broken. Different symmetries will be applicable depending on the exact layout of the surface code.

\subsection{Including Translation Symmetry by using Centered Data}

We start by describing the concepts using the example of translation invariance. Later it will be described how to incorporate exchange invariance and other symmetries.

It is expected that two syndromes that only differ by a translation should have the same logical error. (Some care has to be taken here because it is implicitly assumed that the underlying decoder respects the translation invariance, more on this below.) With infinite training data an \ac{hld} can learn this invariance by "brute-force". Because generating training data is experimentally expensive, it is better to explicitly include this invariance. We can define the \textit{translation class} of a syndrome $s$ as the set of all syndromes that differ from $s$ only by a translation. To explicitly include translation invariance in the training of an \ac{hld}, one unique syndrome in each translation class of syndromes is defined as its \textit{translation representative}. The training data is then pre-processed by mapping each syndrome to its translation representative. Of course, when decoding the syndromes they also need to be pre-processed. This costs some additional computational resources during decoding, that are estimated in \autoref{sec:EstimateScaling}. The pre-processing guarantees that the \ac{hld} includes the translation invariance, and thus reduces the amount of different syndromes the decoder needs to learn.

Explicitly, a pre-processing function can be constructed by using a lexicographic order of the syndromes. First, an arbitrary enumeration of the vertices and plaquettes is chosen. Here, we choose the convention to enumerate from top left to bottom right on the lattice, i.e. the top left vertex is the first, the vertex to its right is the second and so on. Using this enumeration, syndromes can be represented as binary vectors. The $i$'th entry of such a vector is 1 if the $i$'th vertex has a detection, and 0 otherwise. Analogously one defines a vector for the plaquette detections. The vector representing the plaquette result is appended to the vector representing the vertex result. We can then define a total order on syndromes as follows:

\begin{definition}[Lexicographic order of syndromes]
\label{Def:LexicographicOrdering}
For two syndromes $s_1, s_2$ represented as binary vectors, define $s_1 < s_2$ if the first non-zero entry of $s_1 - s_2$ is $1$.  
\end{definition}
In other words, $s_1 < s_2$ if the first non-zero entry of $s_1$ comes "before" the first non-zero entry of $s_2$. The subtraction in the definition is NOT meant mod 2. Note that if $s_1 - s_2 = 0$, so no non-zero entries exist in $s_1 - s_2$, then $s_1 = s_2$. It is easy to verify that definition \ref{Def:LexicographicOrdering} defines a total order on syndromes. In the following, the minimum of a set of syndrome is always meant to be the minimum according to the order in definition \ref{Def:LexicographicOrdering}.

Using this order of syndromes and the enumeration of vertices and plaquettes above, a "centering" algorithm that maps a syndrome to a well defined translation representative can be defined as follows:

\begin{algorithmdescription}[Centering]
Given a syndrome $s$, first compute all possible ways to translate it such that the stabilizer measurement represented by the first vertex of the code detects an error. If there are no vertex detections in the syndrome, instead compute the ways to translate it such that the stabilizer measurement represented by the first plaquette of the code detects an error. Then, compare all the translated syndromes according to the order in definition \ref{Def:LexicographicOrdering} and choose the minimal one according to this order.
\end{algorithmdescription}

\begin{algorithm}[H]
    \Input{Syndrome $s$ as binary vector of length $2L^2$}
    \Output{Translation representative $s_c$ of the syndrome as binary vector}

	T $\leftarrow$ $\{$Translation(s) $|$ first element of Translation(s) is 1$\}$ \;
	\If{T not empty}{ $s_c \leftarrow min(T)$ \tcp*{Minimum over syndromes according to def. \ref{Def:LexicographicOrdering}} \; return $s_c$ }
	\Else{
	T $\leftarrow$ $\{$Translation(s) $|$ element $L^2+1$ of Translation(s) is 1$\}$ \;
	\If{T not empty}{ $s_c \leftarrow min(T)$ \tcp*{Minimum over syndromes according to def. \ref{Def:LexicographicOrdering}}  \; return $s_c$}
	\Else{return empty syndrome}
	}

\caption{Centering algorithm used to obtain a unique translation representative for each syndrome.}   
\label{alg:Centering}
\end{algorithm}  
\

\begin{figure}[t]
\captionsetup[subfigure]{justification = centering}
\begin{subfigure}[]{0.3\linewidth}
\includegraphics[width = \textwidth]{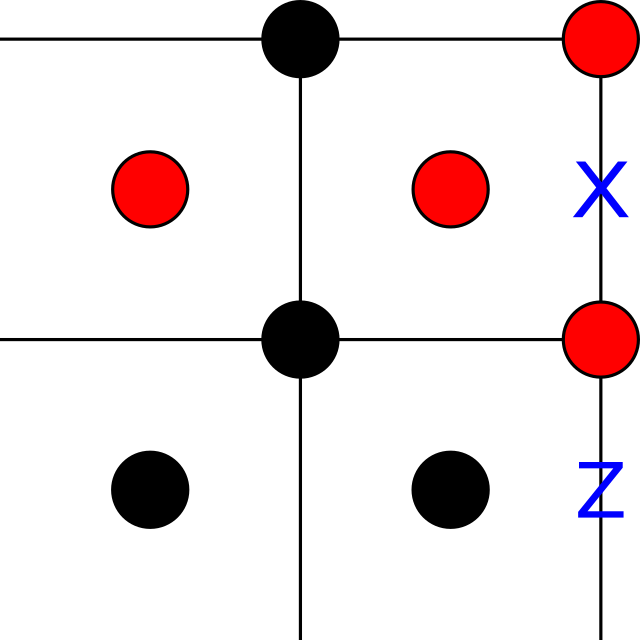}
\caption{}
\label{fig:MWPMTranslationProblem1}
\end{subfigure}
\hfill
\begin{subfigure}[]{0.3\linewidth}
\includegraphics[width=\textwidth]{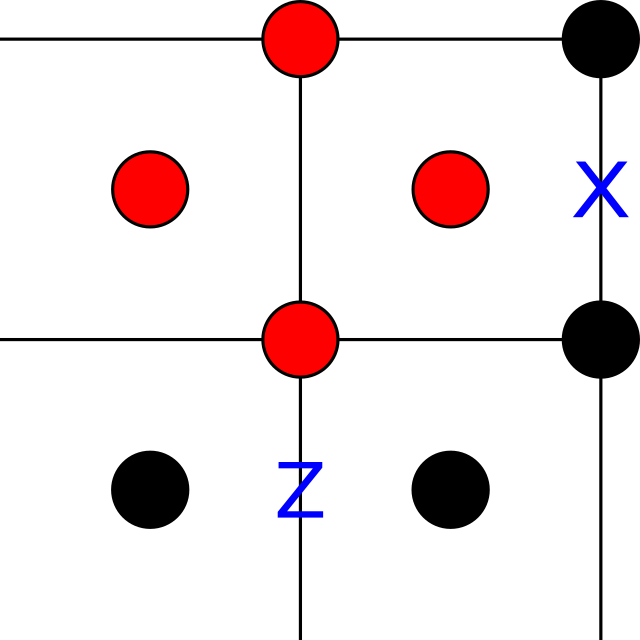}
\caption{}
\label{fig:MWPMTranslationProblem2}
\end{subfigure}
\hfill
\begin{subfigure}[]{0.3\linewidth}
\includegraphics[width=\textwidth]{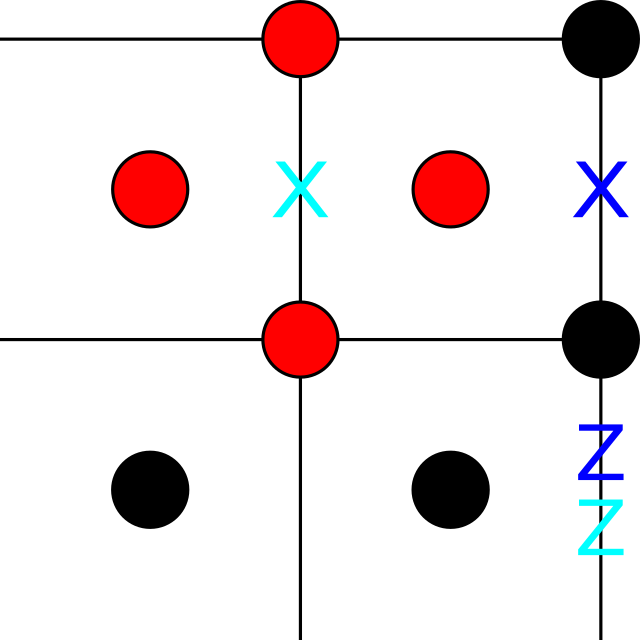}
\caption{}
\label{fig:MWPMTranslationProblem3}
\end{subfigure}
\caption{The red (light gray) dots show two syndromes (\subref{fig:MWPMTranslationProblem1}) and (\subref{fig:MWPMTranslationProblem2}) on a $2\times2$ toric code that differ by a one step translation in the horizontal direction. In blue (dark gray), the recoveries proposed by \ac{mwpm} decoding are shown. (\subref{fig:MWPMTranslationProblem3}) shows the recovery proposed by normal \ac{mwpm} in blue (dark gray). The recovery one obtains by applying \ac{mwpm} to the centered syndrome (\subref{fig:MWPMTranslationProblem2}) and then translating back is shown in bright blue (light gray). Notice that the proposed recoveries differ by a logical $X$ operator.}
\label{fig:MWPMTranslationProblem}
\end{figure}

Since definition \ref{Def:LexicographicOrdering} defines a total order, the minimum according to this order of a list of syndromes is unique. Therefore, algorithm \ref{alg:Centering} will result in a uniquely defined representative of each translation class. A scaling analysis and more details on the implementation of this algorithm can be found in \autoref{sec:EstimateScaling}.
Of course, this algorithm straightforwardly generalizes to any other possible symmetry. In order to find a unique represenant, one first computes all possible representatives and then chooses the minimal one according to the lexicographic order. Finally, it should be noted that the underlying decoder does not necessarily respect the translation invariance of the code. Given two syndromes that only differ by a translation, it is possible that the underlying decoder returns two recoveries that do not only differ by a translation, but by a translation and a logical operator. A simple example of this problem on a $2 \times 2$ code for \ac{mwpm} is shown in \autoref{fig:MWPMTranslationProblem}. Therefore it is important that all syndromes are centered before applying the underlying decoder to them. The recovery proposed by the underlying decoder then needs to be translated back to match the original syndrome. In this way it is guaranteed that the underlying decoder is compatible with translation invariance. The same principle applies to all other invariances one might want to incorporate.

\subsection{Including Exchange Invariance by using Aligned Data}

In addition to translation invariance, which has the largest effect, further symmetries can be included. Here, the case of exchange invariance is considered.
 The basic principle for pre-processing is the same as for translation invariance: One first computes the two possible anti-transpositions of the syndrome, then chooses the one that is minimal according to the lexicographic order (definition \ref{Def:LexicographicOrdering}). Again, pre-processing should take place before applying the underlying decoder, and the proposed recoveries need to be anti-transposed back to match the original syndrome. Furthermore, as mentioned above, an anti-transposition corresponds to a relabeling of the logical operators. The logical $Z_1$ and $Z_2$ operators are exchanged with each other, and the logical $X_1$ and $X_2$ operators are exchanged with each other. Therefore, here, the class labels of the training data need to be adapted if there were anti-transpositions in the pre-processing. Similarly, the logical error proposed by the high level decoder during online decoding needs to be corrected for the effect of anti-transpositions. Note that no such correction was necessary in the case of translation invariance as the logical operators are invariant under translations.

\begin{figure}[t]
\captionsetup[subfigure]{justification = centering}
\begin{subfigure}[]{0.45\linewidth}
\includegraphics[width = \textwidth]{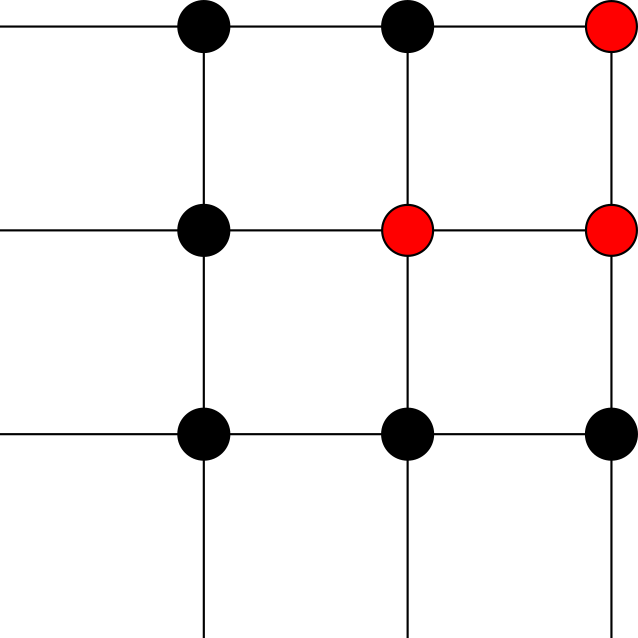}
\caption{}
\label{fig:TranspositionAndTranslationCompatibility1}
\end{subfigure}
\hfill
\begin{subfigure}[]{0.45\linewidth}
\includegraphics[width=\textwidth]{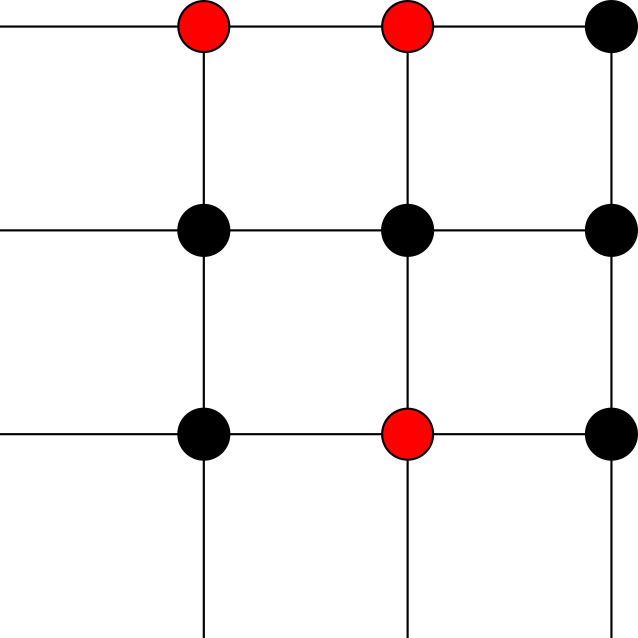}
\caption{}
\label{fig:TranspositionAndTranslationCompatibility2}
\end{subfigure}
\\
\begin{subfigure}[]{0.45\linewidth}
\includegraphics[width=\textwidth]{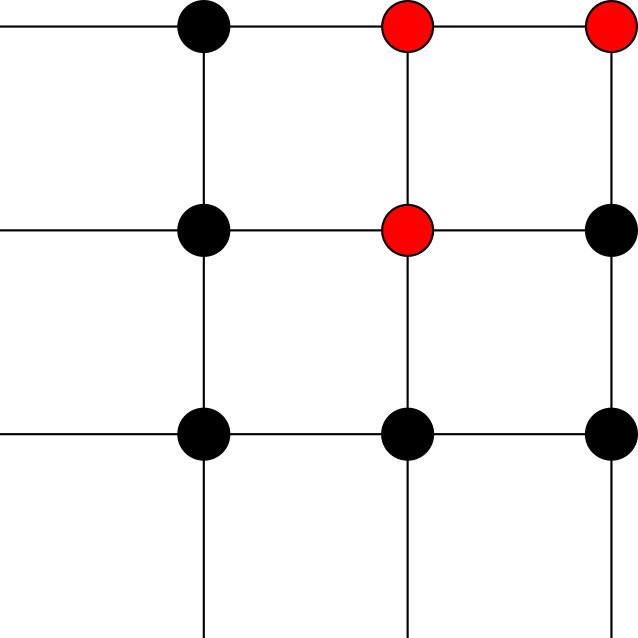}
\caption{}
\label{fig:TranspositionAndTranslationCompatibility3}
\end{subfigure}
\hfill
\begin{subfigure}[]{0.45\linewidth}
\includegraphics[width=\textwidth]{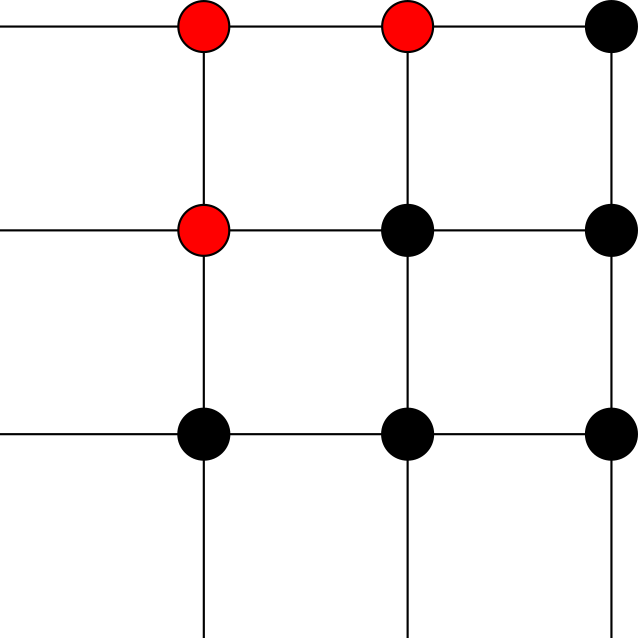}
\caption{}
\label{fig:TranspositionAndTranslationCompatibility4}
\end{subfigure}
\caption{Example of why the naive approach to combining transposition and translation invariance does not work. The syndromes (\subref{fig:TranspositionAndTranslationCompatibility1}) and (\subref{fig:TranspositionAndTranslationCompatibility2}) differ by a translation (one step left and one step to the top). Computing the anti-transposition representative of (\subref{fig:TranspositionAndTranslationCompatibility1}) one obtains syndrome (\subref{fig:TranspositionAndTranslationCompatibility3}) and then centering it one obtains syndrome (\subref{fig:TranspositionAndTranslationCompatibility4}). However the anti-transposition representative of syndrome (\subref{fig:TranspositionAndTranslationCompatibility2}) is syndrome (\subref{fig:TranspositionAndTranslationCompatibility2}) itself, and centering it one obtains again syndrome (\subref{fig:TranspositionAndTranslationCompatibility2}).}
\label{fig:TranspositionAndTranslationCompatibility}
\end{figure}

In the following, we will use $s^{t}$ to denote the anti-transposition representative of a syndrome $s$. 
When combining both translation and exchange invariance, the naive approach is to first compute the representative of the anti-transposition class of a syndrome, and then center this representative. This approach does not work, as illustrated with an example on the $3 \times 3$ toric code in \autoref{fig:TranspositionAndTranslationCompatibility}. The syndrome (\subref{fig:TranspositionAndTranslationCompatibility2}) differs from the syndrome  (\subref{fig:TranspositionAndTranslationCompatibility1}) by a translation one step to the left and one step to the top, taking into account the periodic boundary conditions. Therefore they belong to the same translation class and must be mapped to the same representative. However, if one first computes the anti-transposition representative of the syndrome (\subref{fig:TranspositionAndTranslationCompatibility1}) and then centers it, one obtains the syndrome (\subref{fig:TranspositionAndTranslationCompatibility4}). If one does the same for the syndrome in (\subref{fig:TranspositionAndTranslationCompatibility2}), one obtains the syndrome in (\subref{fig:TranspositionAndTranslationCompatibility2}) itself. This illustrates that the naive approach assigns different representatives to different translations of the same syndrome. Therefore, a slightly more complicated algorithm has to be used to actually compute unique representatives for each class. This algorithm will be called "alignment" algorithm and is described in the following:

\begin{algorithmdescription}[Alignment]
Given a syndrome $s$, it is first centered to obtain $s_c$. Then, the anti-transposition representative $s_c^t$ of $s_c$ is computed and also centered to obtain $(s_c^t)_c$. The two syndromes $s_c$ and $(s_c^t)_c$ are compared and the minimal of the two is chosen. This pre-processing will map syndromes that differ only by translations and anti-transpositions to the same syndrome.
\end{algorithmdescription}

\begin{algorithm}[H]
    \Input{Syndrome $s$ as binary vector of length $2L^2$}
    \Output{representative $s_a$ of the syndrome  under combined translations and anti-transpositions as binary vector}

	$s_c \leftarrow$ Center($s$)\;
	$s_c^t \leftarrow$ Anti-transposition representative($s$)\;
	$(s_c^t)_c \leftarrow$ Center($s_c^t$)\;
	return min($s_c$, $(s_c^t)_c$)  \tcp*{Minimum over syndromes according to def. \ref{Def:LexicographicOrdering}} 
    
\caption{Alignment algorithm used to obtain a unique representative under combined translations and anti-transpositions for each syndrome.}   
\label{alg:Alignment}
\end{algorithm}  
\

Again, the underlying decoder might not be compatible with the alignment by default. To rectify this issue, the same strategy as above is employed. Instead of decoding a syndrome $s$ directly, the aligned syndrome $s_a$ is decoded. All transformations (both translations and anti-transpositions) applied to $s$ in order to obtain $s_a$ are tracked. The recovery proposed by the underlying decoder is then transformed back to match the original syndrome $s$.

\subsection{Estimate Scaling of the Centering Algorithm}
\label{sec:EstimateScaling}

The use of symmetries as described above introduces an additional overhead during the operation of the decoder, since each new syndrome has to be pre-processed, not only during training but also during the online decoding. To get an idea of how significant the overhead introduced by the centering algorithm is, an estimate of the runtime scaling with the code distance is given here. As the alignment algorithm mainly consists of multiple applications of centering, the scaling will be the same.

Because the centering algorithm makes use of the translation of syndromes, we first consider the cost of such an operation. It strongly depends on the data structure that is used to represent syndromes. If a syndrome is simply represented as a binary array and the translation is done by creating a new array with shifted entries it will take linear time in the number of elements, thus $\mathcal{O}(L^2)$. A more efficient representation for our purposes is possible by using 2-dimensional instead of 1-dimensional arrays as follows:  A vector of vertex detections can be represented as a 2-dimensional array. Each row of the array represents one row of the code (with the usual convention that an entry is $1$ if the corresponding vertex has a detection and $0$ otherwise). Iterating over a syndrome can then be done by iterating row-wise over the array, possibly taking into account periodic boundary conditions. 
Translating the vertex detections can then be done without copying by simply changing the starting index of the iteration. 
Of course, the same iteration method applies to the plaquette detections. Therefore translation of a syndrome is an $\mathcal{O}(1)$ operation in this representation. 

The centering algorithm, as described in \autoref{alg:Centering}, then takes as input the measured syndrome as a vector of vertex and plaquette detections, each of length $L^2$ and represented as a 2-dimensional array as described above. The first step of the algorithm is to find all non-zero entries in the vector of vertex detections. Then one translation is computed for each non-zero entry (each detection), corresponding to shifting this entry to the first place in the vector. Then, these different translations are compared according to the lexicographic order (definition \ref{Def:LexicographicOrdering}) and the minimal one is returned. If there were no non-zero entries in the vector of vertex detections, the same procedure is done instead for the plaquette detections.

Finding all nonzero entries of the input vector takes time $\mathcal{O}(L^2)$. Computing one translation of the syndrome is $\mathcal{O}(1)$ as described above. The input vector has one non-zero entry for each detection in the syndrome. Because each single qubit error on the toric code creates at most two detections (less if there are neighbouring errors), the average amount of detections is at worst proportional to the average amount of errors, which is $pL^2$ if errors happen at rate $p$. Therefore, on average, $\mathcal{O}(L^2)$ different translations must be computed in the centering algorithm. Then, the minimal one of these according to definition \ref{Def:LexicographicOrdering} has to be found. Finding the minimum of a list with $n$ elements can be done in $n-1$ comparisons. Each comparison will, in the worst case, take time proportional to the number of elements in the vector, thus $\mathcal{O}(L^2)$. This results in a scaling of $\mathcal{O}(L^4)$ in total. However generally, the comparison will already terminate after comparing the first few elements of two different translations. More precisely, we consider the probability that the $n$'th elements of the two different translations of the syndrome are equal, given that all previous elements were equal. Because only neighboring vertices are correlated, this probability can be upper bounded by some value $p_e < 1$ independent of the code size. The probability that the comparison terminates after $n$ steps is then upper bounded by:
\begin{equation}
p_n = p_e^{n-1}*(1-p_e)
\end{equation}
Therefore the comparison will terminate after an average amount of steps that is upper bounded by: 
\begin{equation}
 \overline{n} = \sum_{n=0}^{L^2} np_e^{n-1}(1-p_e) 
\end{equation}
This can be further upper bounded by a constant independent of $L$:
\begin{equation}
 \overline{n} = \sum_{n=0}^{L^2} np_e^{n-1}(1-p_e) < \sum_{n = 0}^{\infty} np_e^{n-1}(1-p_e) = \frac{1}{1-p_e}
\end{equation}

Therefore the comparison will be $\mathcal{O}(1)$ in the average case. This gives a total average case complexity of $\mathcal{O}(L^2)$.

As a point of reference, standard \ac{mwpm} decoding has a scaling of $\mathcal{O}(L^6)$. It can be optimized to scale as $\mathcal{O}(L^2)$, and can be parallelized to achieve $\mathcal{O}(1)$ scaling \cite{FowlerImprovedMWPM}. The trivial decoder described in \autoref{sec:SimpleDecoders} matches the detections iteratively, so it scales linearly in the number of detections and thus quadratically in $L$. Therefore, the average case scaling of centering matches the scaling of the trivial decoder. Actual values of the execution time will of course be strongly implementation and hardware dependent. On our setup, using an unoptimized Python implementation of the algorithms, the generation of a training data set of $10^4$ errors for an \ac{hld} on a $5 \times 5$ toric code using \ac{mwpm} as the underlying decoder took about 52 $s$. Aligning this data set as described in \autoref{alg:Alignment}, both accounting for translation and transposition invariance, took about 11 $s$. Both operations were done on an Intel Core i5-8400 CPU. (Only a single core was used.)  In conclusion, there is hope that the additional overhead during decoding that arises from the inclusion of symmetries is manageable.
 
Another important question is how much the centering will reduce the amount of training data that is needed to train good decoders. Unfortunately, it is difficult to give rigorous estimates here. In general, the number of different syndromes the decoder needs to classify correctly is exponential in the code size. Thus one might expect that the amount of training data needed is also exponential in the code size. For each syndrome, there are $L^2$ different translations of this syndrome. In the limit of small error rates, where there are few detections in each syndrome, only few translations are identical. Then the reduction in the amount of training data achieved by centering is at best of order $L^2$. Therefore centering is not sufficient to combat the exponential scaling of the amount of training data. However, it does allow for very noticeable improvements if the amount of training data is not too small, as will be seen in  \autoref{sec:NumericalResults}. As a short summary of the results presented there: On our setup, it was possible to train good decoders for toric codes of up to size $5 \times 5$ without using symmetries, but the inclusion of symmetries offered noticeable advantages in decoding accuracy. On the $7 \times 7$ training a good decoder was only possible when using symmetries. Training a good decoder for a $9 \times 9$ code was not possible even when using $10^8$ training data points and exploiting symmetries. 

\section{Numerical Results}
\label{sec:NumericalResults}

The algorithms described above were tested on the $3 \times 3$, $5 \times 5$ and $7 \times 7$ toric code. Different \acp{ffnn} were trained for use in \aclp{hld}. Networks were trained incorporating either no symmetries (uncentered data), only translation symmetry (centered data) or both translation and exchange symmetry (aligned data). As a shorthand, networks trained with uncentered / centered / aligned data are sometimes refered to as uncentered / centered / aligned networks. For simplicity, the training data was always generated at noise parameter $p = 0.01$. The weights of the networks were always initialized from a normal distribution with width $0.01$. For stochastic gradient descent, a batch size of $1000$ was used. No weight decay was employed unless otherwise specified. Following \cite{Varsamopoulos_2017}, two hidden layers with decreasing number of units were used. The input layer had the size $2L^2$, corresponding to the size of a syndrome, and the output layer had the size 16, corresponding to the 16 possible logical errors. Note that this is in contrast to the decoders tested in \cite{Varsamopoulos_2017, Chamberland_2018, VarsamopoulosDistributedDecoders} on the rotated surface code, where only 4 logical errors were possible. Therefore this work also shows that the high level decoding scheme can be applied to surface codes with a larger number of logical qubits. During training, the performance of the decoder was monitored on a validation set that was generated independently from the training data. This was used to tune the hyperparameters of the network. Furthermore, comparing the \emph{training error} (error on the training set) and the \emph{validation error} (error on the validation set) can be used to see whether the network is overfitting. The trained decoders were tested for depolarizing noise parameters $p = 0.01$ to $p = 0.18$ in steps of $0.01$, resulting in the \emph{test error}. Unless otherwise specified a test set of size $10^6$ was used for each noise parameter. The test sets were generated independently from both training and validation sets to avoid overestimating the performance of the decoder, since training and hyperparameter selection tunes the decoder to the specific training and validation sets \cite[Chap. 5]{Goodfellow}. Error bars in the plots represent $95 \%$ confidence intervals. They were obtained by approximating the logarithm of the ratio of binomial proportions by a normal distribution as described in \cite{BinomialRatioConfidence}. It should be noted that it is also possible to use more layers to achieve slightly higher decoding accuracy at the cost of longer training and execution times. Using three layers, the relative improvement in decoding accuracy at $p = 0.1$ was of order $5 \%$ compared to two layers, both with and without the use of symmetries.  

\begin{figure}[t]
\begin{subfigure}{\linewidth}
\includegraphics[width = \linewidth]{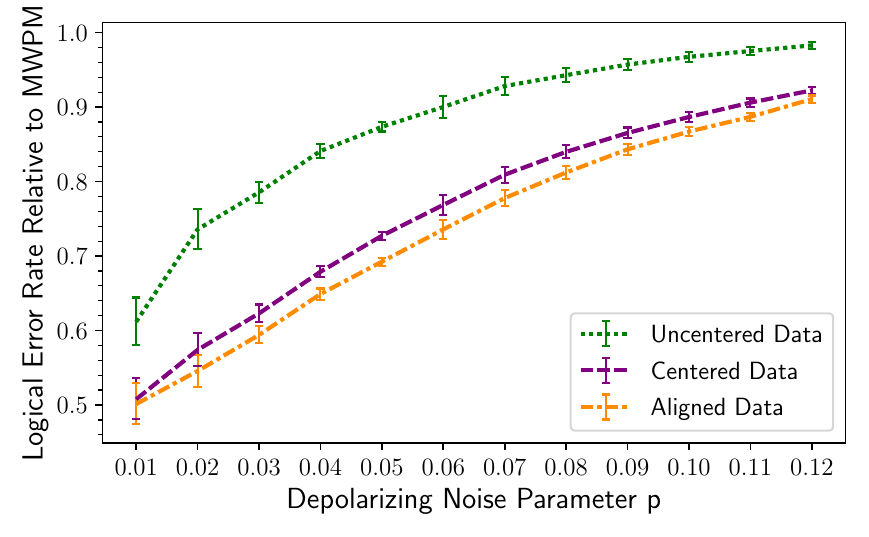}
\caption{} 
\label{fig:HLD5x510MillionRelative}
\end{subfigure}
\begin{subfigure}{\linewidth}
\includegraphics[width = \linewidth]{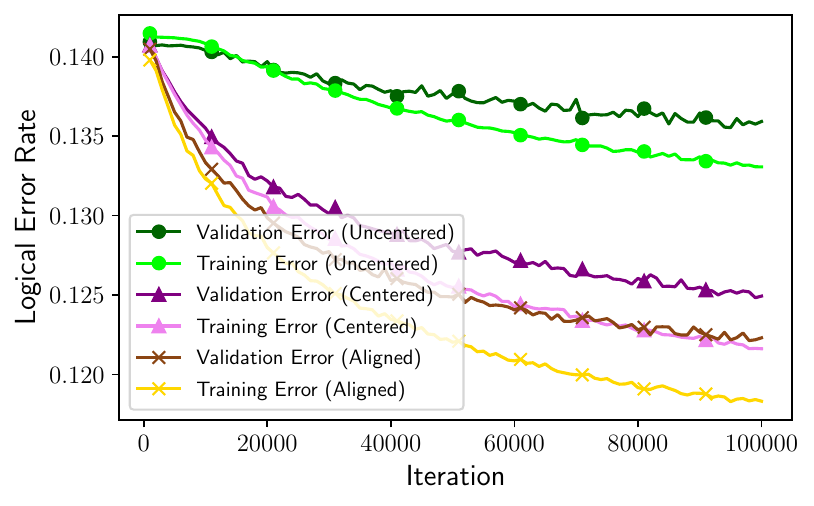}
\caption{} 
\label{fig:HLD5x510MillionTraining}
\end{subfigure}
\caption{Comparison of decoders on the $5\times5$ toric code using aligned, centered or uncentered training data The training data sets had a size of $9*10^6$ and were generated at depolarizing noise parameter $p = 0.1$. The hyperparameters were the same for all networks: layer sizes = $500, 250$, $n_{it} = 10^5$, $\eta = 0.001$. In (\subref{fig:HLD5x510MillionRelative}), plotted is $p_{decoder} / p_{MWPM}$. The dotted lines are only to guide the eye and do not represent actual data points. For $p \leq 0.05$, larger test sets of size $10^7$ ($5*10^7$ for $p = 0.01$) were used to obtain more accurate values for the error rates, which explains the smaller error bar at $p = 0.05$ compared to $p = 0.06$. In (\subref{fig:HLD5x510MillionTraining}) the training of the networks is compared.}
\label{fig:HLD5x510Million}
\end{figure}

We start by considering \acp{hld} on the $5 \times 5$ toric code using \ac{mwpm} as the underlying decoder. Here, using a training set of size $9*10^6$ was sufficient to obtain significant improvements over standard \ac{mwpm} even when not accounting for symmetries. However, when accounting for symmetries, about another $20 \%$ relative improvement could be obtained. The error rates with and without symmetries, relative to \ac{mwpm}, are compared in \autoref{fig:HLD5x510MillionRelative}. Shown is the relative logical error rate $p_{decoder} / p_{MWPM}$ for high level decoders trained on the same dataset, but either not accounting for symmetries (uncentered), accounting only for translation invariance (centered) or accounting for both translation and exchange invariance (aligned). The logical error rate is shown for different depolarizing noise parameters. It can be seen that using translation invariance allows for a large improvement over standard \ac{mwpm}, and further accounting for exchange invariance leads to another small improvement. It is indeed expected that translation invariance leads to larger improvements than exchange invariance. The reasoning is that the translation class of a syndrome contains up to $L^2$ elements, while the anti-transposition class of a syndrome contains only up to 2 elements. The difference between aligned and centered data becomes less pronounced for smaller error rates, as the decoder is more accurate for small syndromes by default.
Considering the training of the decoders, one observes that including symmetries leads to improvements in both validation and training error. (\autoref{fig:HLD5x510MillionTraining}) Therefore the pre-processing actually allows for a more accurate fit even to the training data, i.e. the data was presented in a form more suitable to the model.

\begin{figure}[t]
\includegraphics[width=\linewidth]{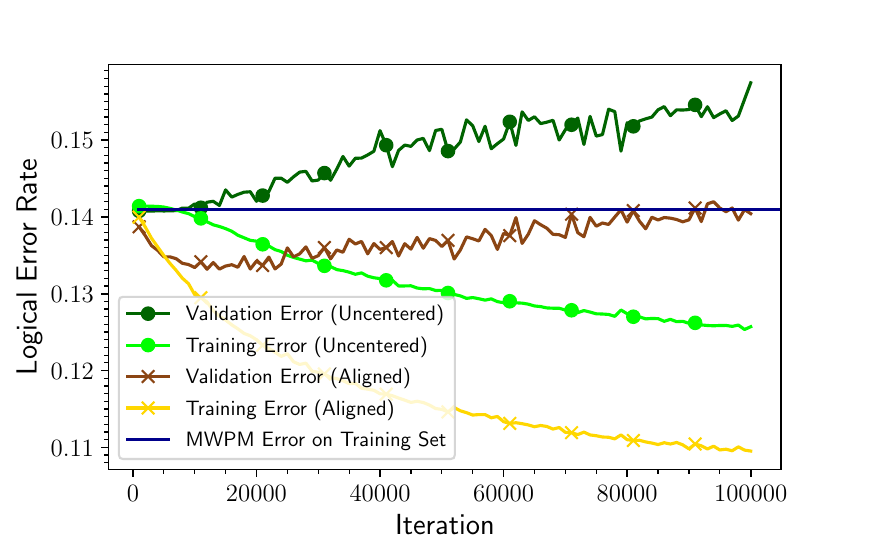}
\caption{Comparison of network training on the $5\times5$ toric code with aligned or uncentered data sets of size $1.8*10^6$, generated at depolarizing noise parameter $p = 0.1$. The small training set size was chosen to test the minimum amount of training data that is needed to obtain improvements in the logical error rate. Network parameters: layer sizes = $500, 250$, $n_{it}$ = $10^5$, $\eta = 0.001$}
\label{fig:FFNN5x5CenteredVSUncentered-2Mil-TrainingComparison}
\end{figure}

To investigate by how much we can reduce of the size of the training set, decoders with uncentered or aligned data were trained with training sets of size $4.5*10^6$, $2.7*10^6$ and $1.8*10^6$. The training for the smallest data size is compared in \autoref{fig:FFNN5x5CenteredVSUncentered-2Mil-TrainingComparison}.
No improvement over \ac{mwpm} could be reached without the use of symmetries. Aligning the data on the other hand does allow for improvements. Again, both validation and training error are improved. However, the validation error does increase again in later iterations of training. Therefore we expect that it is not possible to use even less training data and still obtain good results. The aligned network actually outperformed \ac{mwpm} for all tested error rates up to the pseudo-threshold of around $0.12$. (The pseudo-threshold is the noise parameter at which the logical error rate matches the error rate of two unencoded qubits.) Using $2.7*10^6$ data points, the uncentered network started to outperform \ac{mwpm}, but only for error rates $p < 0.05$. Consistent improvements using uncentered data were only reached using $4.5*10^6$ training data points. This clearly shows that the size of the training set can be noticeably reduced when employing symmetries. Furthermore, we can compare the validation errors (there was no significant difference between validation and test error here) in \autoref{fig:FFNN5x5CenteredVSUncentered-2Mil-TrainingComparison} (small training set) and \autoref{fig:HLD5x510MillionTraining} (large training set).
Stopping after about 10000 iterations to avoid overfitting, the small training set with alignment could be used to achieve a validation error of about 0.135, while the training on the large set but without symmetries lead to a validation error slightly above 0.135. Thus, with symmetries significantly less data is required to achieve the same performance compared to the case without symmetries.  

\begin{figure}[t]
\includegraphics[width = \linewidth]{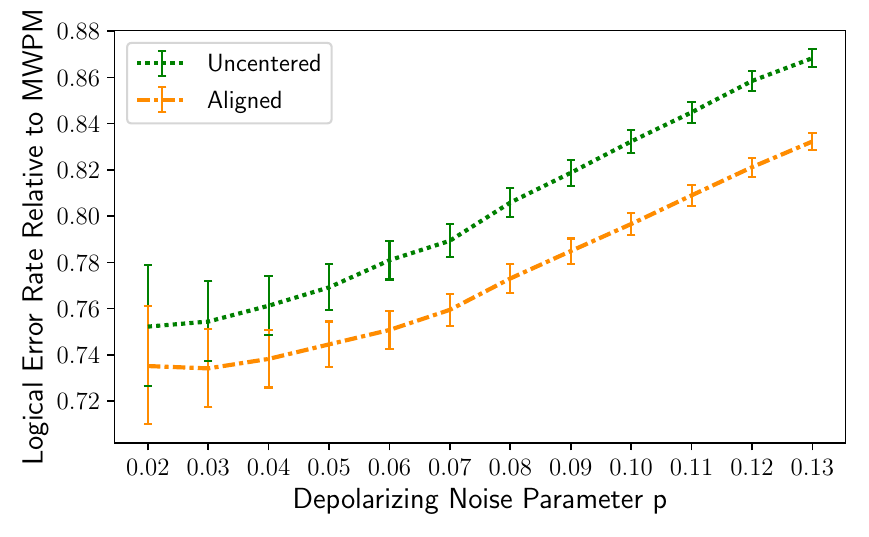}
\caption{Performance of a high level decoder on the $3\times3$ toric code using an aligned or uncentered training data set of size $9*10^5$ generated at depolarizing noise parameter $p = 0.1$. Plotted is $p_{decoder} / p_{MWPM}$. Network parameters: layer sizes = $500, 250$, $n_{it} = 10^5$, $\eta = 0.001$}
\label{fig:FFNN3x3-RelativeError}
\end{figure}

Similar effects could be observed on both the $3\times3$ and the $7\times7$ toric code. On the $3\times3$ code, the improvements gained by employing symmetries are much smaller (\autoref{fig:FFNN3x3-RelativeError}). Here, a data set of size $10^6$ was already sufficient to obtain large improvements over \ac{mwpm} even without the use of symmetries. Two networks were trained, one with aligned data and one with uncentered data. Both networks used 2 hidden layers of sizes $500,250$, a training duration of $10^5$ iterations and a constant learning rate of $0.001$. The relative improvement of the aligned over the uncentered network at $p = 0.1$ was about $4\%$, and at $p = 0.03$ it was about $2\%$. We also considered the training error of the decoder (not shown in the figures). The training error of the aligned decoder was worse than the training error of the uncentered decoder, while the test error was improved as explained above. This is in contrast to the examples on the $5 \times 5$ toric code, where both training and test errors were improved. Therefore it seems that in this case, the inclusion of symmetries mainly helps with generalization and prevents over-fitting to the training data. On the $5 \times 5$ toric code the symmetries were also useful in finding a good fit to the training data at all.
The main reason why the explicit inclusion of symmetries is less important for the $3\times3$ code is that the training set is large enough to learn the invariances by "brute force".
For the $3\times3$ toric code, there are $2^{18} = 262144$ different syndromes, so one expects a large fraction of the possible syndrome to appear in a training set of size $10^6$. However for the $5\times5$ toric code there are $2^{50} \approx 1.1*10^{15}$ different syndromes, so even a training set of size $10^7$ will never cover the whole syndrome space. Therefore, for the $5\times5$ code, it is more important to introduce the invariances.

\begin{figure}[t]
\includegraphics[width = \linewidth]{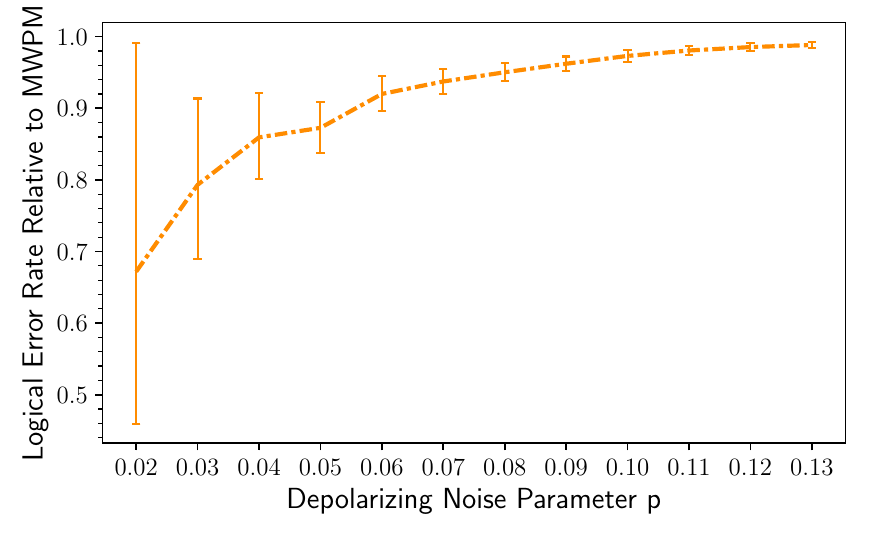}
\caption{Performance of a high level decoder on the $7\times7$ toric code using an aligned training data set of size $4.5*10^7$ generated at depolarizing noise parameter $p = 0.1$. Plotted is $p_{decoder} / p_{MWPM}$. $p = 0.01$ is not plotted as both the decoder and \ac{mwpm} performed perfectly on the test set. The large error bars at low $p$ are due to the very low number of logical errors made by both \ac{mwpm} and the \ac{hld}. Network parameters: layer sizes = $500, 250$, $n_{it} = 10^6$, $\eta = 0.001$}
\label{fig:FFNN7x7Aligned-M4-50Mil-2}
\end{figure}

On the $7\times7$ code decoders were trained using up to $5*10^7$ training examples. Without the use of symmetries, it was not possible to reach any improvements over \ac{mwpm}. However, by aligning the training data, some improvements could be reached. It was possible to slightly outperform normal \ac{mwpm} at all tested error rates.(\autoref{fig:FFNN7x7Aligned-M4-50Mil-2}) The relative improvement was larger for smaller error rates. At large error rates the performance of the decoder was very close to \ac{mwpm}. This is expected, as the larger error rates were close to the pseudo-threshold of the code.

\begin{figure}[t]
\includegraphics[width=\linewidth]{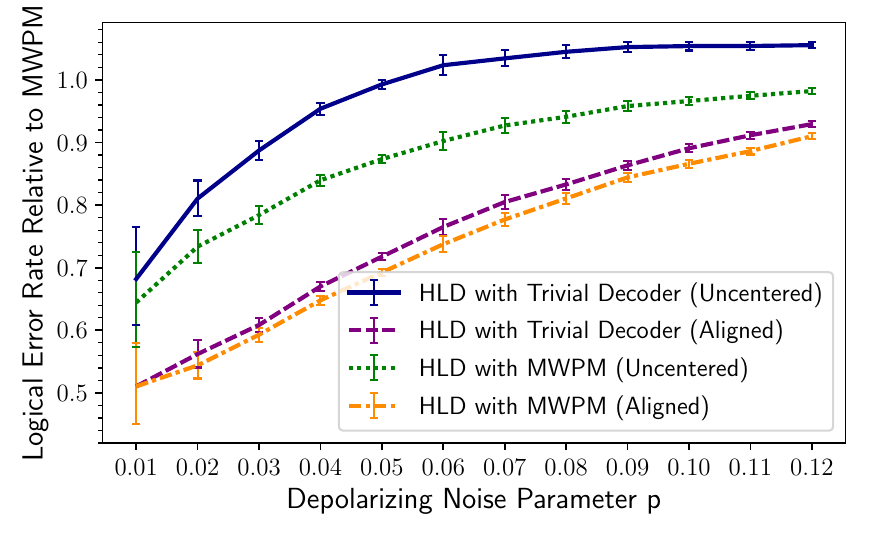}
\caption{Comparison of different high level decoders on the $5\times5$ toric code relative to \ac{mwpm}. Plotted is $p_{decoder} / p_{MWPM}$ for different decoders. Shown are high level decoders based on \ac{mwpm} and high level decoders based on a trivial underlying decoder. In both cases both aligned and uncentered training sets of size $9*10^6$ were used. For $p \leq 0.05$ larger test sets of size $10^7$ were used for increased accuracy.}
\label{fig:5x5FastDecoderRelative}
\end{figure}

As mentioned above, it is also possible to train an \ac{hld} on top of a trivial decoder instead of \ac{mwpm}. This has the advantage that the decoding will be faster, and also training data can be generated faster. Therefore, it was also tested how symmetries affect the performance of an \ac{hld} when using the trivial decoder explained in \autoref{sec:SimpleDecoders} as the underlying decoder. The trivial decoder itself has very bad error rates, worse than those of two unencoded qubits. However the high level decoders based on it still produce good results. From here on we refer to high level decoders based on the trivial decoder as \acs{hldt} \acused{hldt}, and to high level decoders based on \ac{mwpm} as \acs{hldm} \acused{hldm}. Two \acp{hldt} were trained on the $5\times5$ toric code based on the same $10^7$ physical errors also used above for \autoref{fig:HLD5x510MillionRelative}. For one decoder the data was aligned, and the other used uncentered data. These two decoders were compared to the two \acp{hldm} presented in \autoref{fig:HLD5x510MillionRelative}. The same hyperparameters were used for training, with the exception of the training duration, which was longer for the \acp{hldt}. Longer training was necessary for the error rates to converge. The \acp{hldt} were trained for $10^6$ iterations as opposed to $10^5$ iterations for the \acp{hldm}. The comparison of the logical error rates is shown in \autoref{fig:5x5FastDecoderRelative}. The logical error rates are again given relative to standard \ac{mwpm}.
It can be seen that the \acp{hldt} perform worse than the corresponding \acp{hldm}. However, the difference is noticeably smaller when employing symmetries. Furthermore, without symmetries, the \ac{hldt} outperforms \ac{mwpm} only for small depolarizing noise parameters below $0.05$, while for larger noise parameters it performs worse than \ac{mwpm}.
The performance of the \ac{hldt} is noticeably improved by the introduction of symmetries. It outperforms \ac{mwpm} at all noise parameters. In fact, the aligned \ac{hldt} performs better than the uncentered \ac{hldm}. In conclusion, it possible to use a fast but inaccurate underlying decoder to speed up the decoding process. The inclusion of symmetries is especially important in this case to minimize the decrease in accuracy.

\section{Conclusion}

The main result of this paper is that the performance of neural network based decoders for surface codes can be significantly improved by taking into account the symmetries of the code. A pre-processing algorithm with manageable overhead was proposed. This method was tested numerically for the high level neural network based decoder described in \cite{Varsamopoulos_2017}. Tests were done for lattice lengths $L =  3, 5$ and $7$. Significant improvements were observed when accounting for symmetries. This allows for a reduced amount of training data, adressing one of the main problems pointed out in \cite{Chamberland_2018}. It is therefore one step in the direction of scalable neural network based decoders, although it does not seem sufficient by itself. For example, while the use of symmetries allowed for a decoder on the $7\times7$ code, a good decoder on the $9\times9$ code could not be trained with the simple feed-forward architecture considere here, even when using symmetries. Our method of pre-processing should be used to supplement other approaches, such as the use of sophisticated network architectures proposed in  \cite{XiaotongNNDecodersForLargeToricCodes}. In previous work \cite{Chamberland_2018, Varsamopoulos_2017, Varsamopoulos_DesigningNNBasedDecoders} on high level decoders, the underlying decoder was always chosen to be fast but inaccurate. Here, it was experimentally demonstrated that an accurate underlying decoder also leads to a more accurate \acl{hld} in practice, i.e. not assuming perfect training. However, it was also shown that an inaccurate underlying decoder can still lead to good results if the training is good enough. Therefore, the improvements reached by including symmetries were especially important in the case of a fast underlying decoder, which is also the most interesting case in practice. Additionally, it was shown that neural network based decoders can be applied to surface codes encoding more than one qubit. Although the inclusion of symmetries was demonstrated here for high level decoders, the core ideas and the pre-processing algorithm can likely be applied to other decoders. 

For future research, it would be interesting to further test the methods presented here on more realistic noise models, especially with imperfect syndrome extraction, and for different network architectures like convolutional and recurrent neural networks that have been shown to outperform simple feed forward neural networks \cite{Varsamopoulos_DesigningNNBasedDecoders}. It would also be interesting to test these methods for low level decoders (e.g.  \cite{NeuralDecoder}). Furthermore, as mentioned above, the use of symmetries alone does not seem sufficient to allow for scalable neural network based decoders. Therefore it would be interesting to combine this approach with decoders based on local decompositions of the code (e.g. \cite{XiaotongNNDecodersForLargeToricCodes} and \cite{VarsamopoulosDistributedDecoders}).

\begin{acronym}
\acro{ffnn}[FFNN]{feed forward neural network}
\acro{relu}[ReLU]{rectified linear unit}
\acro{iid}[i.i.d.]{independent and identically distributed}
\acro{rbm}[RBM]{restriced boltzmann machine}
\acro{mld}[MLD]{maximum likelihood decoding}
\acro{mwpm}[MWPM]{minimum weight perfect matching}
\acro{lld}[LLD]{low level decoder}
\acro{hld}[HLD]{high level decoder}
\acro{hldm}[HLDM]{high level decoder based on MWPM}
\acro{hldt}[HLDT]{high level decoder based on a trivial decoder}
\end{acronym}

\begin{acknowledgments}
We thank Kai Meinerz for interesting discussions about surface code decoding with neural networks. This project was funded by the Deutsche Forschungsgemeinschaft (DFG, German Research 
Foundation) under Germany's Excellence Strategy – Cluster of Excellence 
Matter and Light for Quantum Computing (ML4Q) EXC 2004/1 – 390534769
\end{acknowledgments}

\bibliography{Bibliography}

\end{document}